\pdfoutput=1
\documentclass[pra,onecolumn,showpacs,floatfix,superscriptaddress,nofootinbib]{revtex4}
\usepackage{amsfonts,palatino,amsthm,latexsym}
\usepackage[fleqn]{amsmath}
\usepackage{amssymb}
\usepackage{epsfig}
\usepackage{epstopdf}
\usepackage{textcomp}
\usepackage{cancel}
\usepackage{wrapfig}
\usepackage{bbm}
\usepackage[normalem]{ulem}
\usepackage{multirow}
\def\eali\end{align*}
\def\be{\begin{equation}}
\def\ee{\end{equation}}
\def\bea{\begin{eqnarray}}
\def\eea{\end{eqnarray}}
\def\bal{\begin{align}}
\newcommand{\eal}{\end{align}}
\def\ble{\begin{flalign}}
\newcommand{\ele}{\end{flalign}}
\def\ba{\begin{array}}
\def\ea{\end{array}}
\def\bali{\begin{align*}}
\def\nn{\nonumber \\}

\def\bite{\begin{itemize}}

\def\eat{\end{itemize}}
\def\begfig{\begin{figure}[h!]}
\def\endfig{\end{figure}}

\newcommand{\bwe}{\begin{widetext} \begin{eqnarray}}
\newcommand{\ewe}{\end{eqnarray}\end{widetext}}

\def\ox{\otimes}
\def\dag{\dagger}

\def\ket#1{\mid #1 {\cal{i}}}

\def\abs#1{\mid #1 \mid}
\def\com#1#2{[#1,#2]}
\def\norm#1{\| #1 \|}

\def\imp{\Rightarrow}

\def\Z{\ensuremath{\mathbb{Z}}}

\def\R{\ensuremath{\mathbb{R}}}

\def\rf#1{Eq.(\ref{#1})}

\def\bd{\bf \begin{definition}: \it}
\def\ed{\end{definition} \rm}
\def\blem{\bf \begin{lemma}: \it}
\def\elem{\end{lemma} \rm}
\def\bthe{\bf \begin{theorem}: \it}
\def\ethe{\end{theorem} \rm}
\def\bcor{\bf \begin{corollary}: \it}
\def\ecor{\end{corollary} \rm}
\def\bpro{\bf \begin{proof}: \rm}
\def\epro{\end{proof} \rm}

\def\totalback{\!\!\!\!\!\!\!\!\!\!\!\!\!\!\!\!}
\def\halfback{\!\!\!\!\!\!\!}
\def\undback16{& \!\!\!\!\!\!\!\!\!\!\!\!\!\!\!\!}
\def\su2{\ensuremath{\mathfrak{su(2)}}}
\def\SU2{\ensuremath{{SU(2)}}}

\def\+{\ensuremath{\ket{+}}}
\def\-{\ensuremath{\ket{-}}}

\def\Hs{\ensuremath{\mathcal{H}}}

\def\nnn{\ensuremath{\overline{n}}}
\def\DDD{\ensuremath{\overline{D}}}

\begin{document}
\title{Lieb-Robinson bounds with dependence on interaction strengths}

\author{Isabeau Pr\'emont-Schwarz}
\email{ipremont-schwarz@perimeterinstitute.ca}
\affiliation{Perimeter Institute for Theoretical Physics\\
31 Caroline St. N, N2L 2Y5, Waterloo ON, Canada}
\affiliation{Department of Physics, University of Waterloo \\ Waterloo, Ontario N2L 3G1, Canada}
\author{Jeff Hnybida}
\email{jhnybida@perimeterinstitute.ca}
\affiliation{Perimeter Institute for Theoretical Physics\\
31 Caroline St. N, N2L 2Y5, Waterloo ON, Canada}
\affiliation{Department of Physics, University of Waterloo \\ Waterloo, Ontario N2L 3G1, Canada}

\begin{abstract}
We propose new Lieb-Robinson bounds (bounds on the speed of propagation of information in quantum systems) with an explicit dependence on the interaction strengths of the Hamiltonian.  For systems with more than two interactions it is found that the Lieb-Robinson speed is not always algebraic in the interaction strengths.  We consider Hamiltonians with any finite number of bounded operators and also a certain class of unbounded operators.  We obtain bounds and propagation speeds for quantum systems on lattices and also general graphs possessing a kind of homogeneity and isotropy. One area for which this formalism could be useful is the study of quantum phase transitions which occur when interactions strengths are varied.
\end{abstract}

\maketitle
\section*{Introduction} \label{sec_intro}
In quantum field theory, the principle of locality is enforced by an exact light cone. Whenever two (bosonic) observables are spacelike separated they have to commute, so that neither can have any causal influence on the other. In nonrelativistic quantum mechanics, no explicit request for locality is imposed and it is possible in principle to signal between arbitrarily distant points in an arbitrarily short time, as can be seen from simple perturbative arguments. Nevertheless, the simple perturbation analysis shows that such an influence must decay exponentially with the distance between the observables. The seminal work by Lieb and Robinson \cite{lieb} has made the preceding statement rigorous. In essence, it states that any quantum system whose Hilbert space is composed of a tensor product of local Hilbert spaces and whose Hamiltonian is the sum of local operators will have an effective maximum speed of signals. 
The lightcone is only approximate since it possesses an exponentially decaying tail.

Recently, the Lieb-Robinson bounds have received renewed interest and have been applied to many areas in theoretical condensed matter and quantum information theory. In particular, they were used to prove that gapped systems have exponential clustering in their ground state \cite{hastingsa, clustering}.  The structure of correlations in quantum lattice systems is of particular importance in condensed-matter theory and quantum information theory \cite{Bravyi:2006zz,cramer,Eisert:2006zz,schuch,sims,hastings}. In \cite{Eisert:2006zz}, it is shown that the creation of entanglement can proceed with a speed that is limited by the Lieb-Robinson speed and it is shown that this is useful for proving general scaling laws for entanglement. In \cite{Bravyi:2006zz} these techniques have been exploited to characterise the creation of topological order. Other developments of significant interest include \cite{boso,plenio}, where it is shown that it is possible to entangle macroscopically separated nano-electromechanical oscillators in an oscillator chain and that the resulting entanglement is robust to decoherence. To do this, the coupling between the oscillators is first turned off and the system is brought to its ground state. Once in the ground state, the coupling between the oscillators is turned back on non-adiabatically so as to produce a squeezed state which propagates entanglement over macroscopic distances at a speed related to the Lieb-Robinson speed. Such a system would be of great interest for its possible application as a quantum channel and as a tool for investigating the boundary between the classical and the quantum world.

A systematic application of this technique can yield results for quantum spin models in two dimensions whose exact solution leads to intractable mathematical complexities. In a domain where the vast majority of results are of computational nature, analytical results are very important.

One area of condensed matter physics in which Lieb-Robinson bounds have yet to leave their mark is quantum phase transitions. Quantum phase transitions \cite{nature, Hertz, Kogut, Sachdev, notunderstood} are phase transitions which occur in the quantum regime when the interaction strengths are varied (as opposed to thermodynamic quantities like temperature and pressure in the classical case). So far, with the exception of \cite{LRB1, LRB2} (in which only two interactions have been considered), the Lieb-Robinson bounds have been independent of the relative interaction strengths. When it comes to quantum phase transitions however, the relative interaction strengths are of the essence. In future work, we would be interested in applying the techniques developed in this paper to Quantum Graphity\cite{Konopka:2008hp,Konopka:2008ds}, a model of quantum gravity where it is believed that space emerges from spacetime via a quantum phase transition.

In this paper, we start by reviewing the Lieb-Robinson bound in Section \ref{sec_LR_bounds}. As a precursor, we show in Section \ref{sec_two_interactions_bounded} how the dependence on the interaction strength can be obtained for a system with two interactions. This was done earlier in \cite{LRB1}, but the derivation is presented here in greater detail.  This is also done in \cite{LRB2} for the cases of unbounded operators. We then apply this formalism to the Ising model which is a system of two interactions. In Section \ref{sec_three_interactions} we go on to generalise the previous results to the case of a finite number of interactions and apply it to the XY-model.

\section{The Lieb-Robinson bound} \label{sec_LR_bounds}
In its simplest form, the Lieb-Robinson bound can be derived in the following way. We suppose that the total Hilbert space is the tensor product of local Hilbert spaces:
\begin{align}
 \mathcal H = \bigotimes_x \ \mathcal H_x .
\end{align}
Each local Hilbert space $\mathcal H_x$ is associated to an edge or vertex of a graph $G$ with vertices having a maximum valency $\nu$. The total graph is thus associated to the total Hilbert space. On this Hilbert space we have a local Hamiltonian which defines the evolution:
\begin{align}
H_{local}=\sum_{X\subset G}\Phi _{X} ,
\end{align}
where the operators $\Phi _{X}$ are R-local.  We define an operator to be R-local if it has support on a Hilbert space with an associated subgraph $X$ of $G$ such that the diameter of $X$ in graph distance is less than $R$ and such that $X\cap Y=\emptyset \imp \com{\Phi _{X}}{\Phi _{Y}}=0$.  The norm of these local operators is also assumed to be less than some positive number $K$. The Lieb-Robinson bound is a bound on the commutator of two local observables separated by a distance $d$ at a time $t$ such as
\begin{align}
f(t):=[O_{P}(t),O_{Q}(0)], \label{eqn_commutator}
\end{align}
where the operators $O_P$ and $O_Q$ have support on the Hilbert spaces associated with the subgraphs $P$ and $Q$ which are separated by a graph distance $d$.  Taking the derivative with respect to $t$ and rearranging using the Jacobi identity, we have that
\begin{align}
f^{\prime }(t)=-i\sum_{X\in Z_{P}}\left( [f(t),\Phi _{X}(t)]+\left[
O_{P}(t),[\Phi _{X}(t),O_{Q}(0)]\right] \right)
\label{minipf_one}
\end{align}
where $Z_{P}$ contains subgraphs of diameter at most $R$ which have a non-empty intersection with $P$. Taking the norm and integrating we obtain
\begin{align}
\| [O_{P}(t),O_{Q}(0)]\| \leq \| [ O_{P},O_{Q}]\| + 2\| O_{P}\| \sum_{X\in Z_{P}}\int_{0}^{|t|} ds \| [\Phi_{X}(s),O_{Q}(0)]\|.
\label{minipf_two}
\end{align}
The preceding equation can then be iterated to give
\begin{align}
\Vert \lbrack O_{P}(t),O_{Q}(0)]\Vert \leq \Vert O_{P}\Vert \Vert O_{Q}\Vert\sum_{n=0}^{\infty }\frac{(2|t|)^{n}}{n!}a_{n}
\end{align}
where
\begin{align}
a_{n}:= \sum_{X_{2}\in Z_{X_1}}\ldots\sum_{X_{n+1}\in Z_{X_n}} \prod_{j=1}^{n} \norm{\Phi _{X_j}}\delta^{X_n}_{Q} \leq K^n l^n e^{\lambda (n R - d)} \label{miniproof_three}
\end{align}
for some positive number $\lambda$ and where we have set $X_1:=P$.  We define $\delta^{X}_{Y} = 1$ if $X\cap Y \neq \emptyset$ and $\delta^{X}_{Y} = 0$ otherwise.  We also define $l:= \max_{X}|Z(X)| < \infty$ for $\text{diameter}(X) \leq R$. A bound on $a_n$ is obtained by bounding $\norm{\Phi _{X_j}}$ by $K$ and noticing that there are at most $l$ terms in each sum such that $\delta^{X_n}_{Q}=0$ and $n R < d$. Using this bound for $a_n$ we obtain a form of the Lieb-Robinson bound
\begin{align}
\Vert  [O_{P}(t),O_{Q}(0)]\Vert \leq \Vert O_{P}\Vert \Vert O_{Q}\Vert \exp\left(-\lambda (d- v |t|) \right) ,
\end{align}
where $v\leq (2 K e^{\lambda R})/\lambda$. The preceding inequality implies that the ``speed of signals" in the system is effectively less than $v$. Anything travelling faster than this speed will be exponentially suppressed.

From the preceding derivation of the Lieb-Robinson bound, two limitations immediately surface. First, passing from \rf{minipf_one} to \rf{minipf_two}, the norm of the local operators is used; hence it is necessary that the local operators composing the Hamiltonian be bounded.  Second, it is evident from using $\norm{\Phi _{X_j}}\leq K$ in \rf{miniproof_three} that in obtaining the Lieb-Robinson bound, one has ignored the interplay between different types of operators (different types of interactions) weakening the bound in the case where not all of the local operators have norm $K$. In other words, the bound can be tightened by classifying the different operators making up the Hamiltonian into sets of ``interactions" where all operators in one set commute with each other and come with the same coupling constant in the Hamiltonian and then keeping track of how they appear in the bound. Though the the first limitation cannot be removed entirely \cite{supersonic}, it was partially removed in \cite{LRB2} for a class of unbounded operators having bounded commutators. The purpose of this paper is to alleviate the second limitation.

Before we proceed, however, we point out a stunning implication of the existence of the Lieb-Robinson bound: A relativistic light-cone structure emerges in the continuum limit of certain local (non-relativistic) quantum systems. More precisely, the continuum limit of any quantum system, for which the Lieb-Robinson bound is applicable, will exhibit a sharp emergent light-cone structure with a \emph{finite} signaling speed. A heuristic argument can be given by replacing graph distances with metric distances in some units. For instance, let $\Delta$ be the unit graph distance in some standard unit such as metres or centimetres. If we now rewrite the Lieb-Robinson bound using this standard unit instead of the graph distance, we get
\begin{align}
\norm{\com{O_P(t)}{O_Q(0)}}  \leq  \tilde{M}\exp \Big(\frac{\lambda}{\Delta} \big( v_{L R} \abs{t} - d(P,Q)\big) \Big),
\label{bamboo}
\end{align}
where the Lieb-Robinson speed, $v_{L R}$, is measured in the standard units of length per unit time and the distance $d(P,Q)$ is measured in the same units of distance. It is now clear that if we take the continuum limit of the theory, $\Delta \rightarrow 0$, \rf{bamboo} will be exactly zero if $d(P,Q)>v_{L R}\abs{t}$. A complete analysis of the emergence of a strict light cone in the continuum limit has been shown in \cite{eisertetal} for general harmonic quantum systems.

The Lieb-Robinson bound allows us to conclude that any discrete R-local quantum theory with a Hamiltonian consisting of uniformly bounded or commutator-bounded operators will have an exact emerging light-cone structure in the continuum limit. Since many theories of quantum gravity are built from discrete structures such as graphs, the Lieb-Robinson bound could provide a natural mechanism for the emergence of relativity in those theories without requiring local Lorentz invariance \emph{a priori}.


\section{Theories of Two Interactions} \label{sec_two_interactions_bounded}
Before taking into account the subtle interplay between a finite number of interaction types in the Hamiltonian, we will derive the Lieb-Robinson bound for a Hamiltonian with two interactions. This gives us the intuition needed to derive the bound for a system with more than two interactions.  We consider R-Local quantum systems with Hamiltonians composed of bounded operators containing only two interactions. We will begin by defining precisely what an R-local quantum system is.

\subsection{R-Local Quantum Systems}
R-local quantum systems are quantum systems with interactions of finite range $R$. More rigorously, let $G$ be a graph (finite or infinite) having a maximum valence $\zeta$\footnote{This ensures that the continuum limit of $G$ be locally compact.  The requirement will be needed so that the sum $\sum_{n=0}^\infty c_n \frac{w^n}{n!}$, which we will encounter later, converges for all real $w$.} such that to every edge $e$ we associate a the Hilbert space $\Hs_e\ $ and to every vertex $v$ we associate a Hilbert space $\Hs_v$. The total Hilbert space of the system is then defined to be
\begin{align}
\Hs_{tot}\equiv \bigotimes_{\mbox{e an edge of G}} \Hs_e  \ \ox \ \ \bigotimes_{\mbox{v a vertex of G}}\Hs_v .
\end{align}

We then take the evolution of the system to be $R$-local for $R$, a natural number. By this we mean that the evolution is governed by a Hamiltonian composed of a sum of operators, each having support on a Hilbert space which is associated to a subgraph of $G$ having a diameter less than $R$ (i.e., the graph distance between any two points of the subgraph for which the operator acts is at most $R$). By association, we also refer to the diameter of an operator as the diameter of its associated subgraph. Also, we use the notation that if $\Phi$ is an operator that does not have an explicit time dependence, then it is considered to be an operator evaluated at time $t=0$, i.e., $\Phi = \Phi(0)$; furthermore, we have $\Phi(t) \equiv e^{i t H}\Phi e^{-i t H}$.

\subsection{General Hamiltonians With Two Bounded Interactions}
A common type of Hamiltonian is one with two coupling constants and two different types of operators. We say a Hamiltonian is a general Hamiltonian with two coupling constants if it is of the following type:
\begin{align}
H \equiv \sum_{i\in S_0} h_0 \Phi_0^i + \sum_{j\in S_1} h_1 \Phi_1^j \label{deuxint},
\end{align}
with the properties that for $a\in \Z_2 = \{0,1\}$ and $(i,j)\in S_a^2$ we have $\com{\Phi_a^i}{\Phi_a^j}=0$ and $\norm{\Phi_a^i}=\norm{\Phi_a^j}=1$.  In other words the operators are normalised and operators of the same type at different locations on the graph commute.  Here, the set $S_a$ is the set of labels of the subgraphs of $G$ for which the operators of type $a$ have support.

Let $\Gamma(q,m)$ be the subgraph associated with the operator $\Phi^{m}_{q}$ and let $(a,b) \in \Z_{2}^{2}$.  We define the function
\begin{align}
K_{a \ b}^{i \ j}(t) \equiv \com{\Phi_a^{i}(t)}{\Phi_b^j}, \label{f}
\end{align}
in analogy with Eq. (\ref{eqn_commutator}).  We will generalise the Lieb-Robinson bound as calculated in \cite{LRB1} to R-local Hamiltonians of bounded operators having two interactions by deriving a bound for $K_{a \ b}^{i \ j}(t)$.  Note that since we are dealing with bounded operators, it is possible to absorb the norm of the operators into the coupling constants and we may thus assume $\norm{\Phi_c^k}=1$. In the following, we will show that if $P$ and $Q$ are two separate regions of the graph which are small, as compared to the typical size of the total graph, and $O_P$ and $O_Q$ are bounded local operators acting on the subgraphs $P$ and $Q$ respectively, then
\begin{align}
  \norm{\com{O_P(t)}{O_Q(0)}}\leq \tilde{C}\exp\left(\xi\big(2\frac{\gamma}{\xi} e\sqrt{h_0 h_1} t - d(P,Q)\big)\right) , \label{LRBBG}
\end{align}
where $\tilde{C}$ is a positive constant depending on $O_P$, $O_Q$, and $H$.  Here $\gamma$ and $\xi$ are positive constants depending on the graph structure, and $d(P,Q)$ is the graph distance between $P$ and $Q$. This substantiates the claim \cite{Bravyi:2006zz} that information cannot propagate at speeds greater than $v_{LR} =2\frac{\gamma}{\xi} e\sqrt{h_0 h_1} $ without suffering exponential suppression of the bandwidth as a function of distance. We will call this limit on the speed of information propagation the Lieb-Robinson speed. We obtain this result through recursion, as described in what follows.

Taking the derivative of \rf{f} with respect to $t$, we get
\begin{align}
({K_{a \ b}^{i_1 \ j}}(t))^\prime & =  -i \com{\com{\Phi_a^{i_1}(t)}{H(t)}}{\Phi_b^j} \nn
      & =  -i h_{a+1} \sum_{i_2\in Z_{i_1}}\com{\com{\Phi_a^{i_1}(t)}{\Phi_{a+1}^{i_2}(t)}}{\Phi_b^j} \label{fp} \\
      & =  -i h_{a+1} \sum_{i_2\in Z_{i_1}}\com{\com{\Phi_a^{i_1}(t)}{\Phi_b^j}}{\Phi_{a+1}^{i_2}(t)} +  \com{\Phi_a^{i_1}(t)}{\com{\Phi_{a+1}^{i_2}(t)}{\Phi_b^j}}\nn
      & =  \com{{K_{a \ b}^{i_1 \ j}}(t)}{\Big(-i h_{a+1} \sum_{i_2\in Z_{i_1}} \Phi_{a+1}^{i_2}(t)\Big)}  +  (-i h_{a+1}) \sum_{i_2\in Z_{i_1}}\com{\Phi_a^{i_1}(t)}{\com{\Phi_{a+1}^{i_2}(t)}{\Phi_b^j}},\label{fpp}
\end{align}
where the Jacobi identity was used to obtain the third equality and for $i\in S_q$ with $q \in \Z_2$, we define $Z_i$ to be the set of $k\in S_{q+1}$ such that $\Gamma(q,i)\cap\Gamma(q+1,k)\neq \emptyset$. Note that if $a \in  \Z_2$ and $a=1$ then $a+1 = 0$.  Thus integrating \rf{fpp} we get
\begin{flalign}
&{K_{a \ b}^{i_1 \ j}}(t)  =  T_2(t)  {K_{a \ b}^{i_1 \ j}}(0) -  i h_{a+1} \int_0^t ds \sum_{i_2\in Z_{i_1}}\com{\Phi_a^{i_1}(s)}{\com{\Phi_{a+1}^{i_2}(s)}{\Phi_b^j}} , \label{fb}
\end{flalign}
where $T_2(t)$ is a unitary evolution given by the unitary matrix $U_2(t)= \exp\big(-i h_{a+1} \sum_{i_2\in Z_{i_1}} \Phi_{a+1}^{i_2}(t)\big)$ with $T_2(t) O \equiv U_2(t)^\dag O U_2(t)$ for any operator $O$. Taking the norm of \rf{fb} we obtain
\begin{align}
 \norm{K_{a \ b}^{i_1 \ j}(t)} & \leq  \norm{K_{a \ b}^{i_1 \ j}(0)} +  2 h_{a+1} \int_0^t ds  \sum_{i_2\in Z_{i_1}}\norm{\Phi_a^{i_1}(s)}\norm{\com{\Phi_{a+1}^{i_2}(s)}{\Phi_b^j}}, \label{fail}\\
            & =   \norm{\com{\Phi_a^{i_1}(t)}{\Phi_b^j}} +  2 h_{a+1} \norm{\Phi_a^{i_1}(0)} \int_0^t ds \sum_{i_2\in Z_{i_1}}\norm{\com{\Phi_{a+1}^{i_2}(s)}{\Phi_b^j}}, \nn
           & \leq  2 \delta^{i_1}_{j} +  2 h_{a+1} \int_0^t ds \sum_{i_2\in Z_{i_1}}\norm{\com{\Phi_{a+1}^{i_2}(s)}{\Phi_b^j}}, \nn
           & =  2 \delta^{i_1}_{j}  +  2 h_{a+1} \int_0^t ds \sum_{i_2\in Z_{i_1}}\norm{K_{{a+1} \ b}^{i_2 \ j}(s) },\label{binduct}
\end{align}
where we define
\begin{align}
\delta_{i}^k = \left\{ \begin{array}{cc} 1 & \mbox{if } \Gamma(a_i,i)\cap \Gamma(a_k,k)\neq \emptyset , \\ 0 & \mbox{otherwise}, \end{array}\right. \label{eqn_delta}
\end{align}
and where the last inequality is obtained by realizing that $\norm{\com{\Phi_c^k}{\Phi_d^l}}\leq 2$ and is necessarily zero if $\Gamma(c,k)$ and $\Gamma(d,l)$ have an empty intersection. Applying \rf{binduct} to itself, we get by induction that
\begin{align}
\norm{K_{a \ b}^{i_1 \ j}(t)} & \leq M \sum_{n=0}^\infty \frac{(2 \abs{t} \sqrt{h_{a} h_{a+1}})^n}{n!}\underbrace{\Big(\sum_{i_{2}\in Z_{i_1}}\ldots\sum_{i_{n+1}\in Z_{i_n}}\delta^{i_n}_j\Big)}_{c_n} , \label{binductive}
\end{align}
where $M=\max\left\{\sqrt{\frac{h_1}{h_0}}, \sqrt{\frac{h_0}{h_1}}\right\}$. Note that $c_n$ counts the number of chains associated to $n$ local operators (and thus $n$ subgraphs $\Gamma(c,k)$) starting with $\Gamma(a,i_1)$ and ending with $\Gamma(b,j)$. In all such chains a $\Gamma(0,k)$ is followed by a $\Gamma(1,l)$ where $\Gamma(0,k)\cap\Gamma(1,l) \neq \emptyset$ and vice versa. If we repeat exactly the same procedure, but this time with $O_P(t)$ instead of $\Phi_a^{i_1}(t)$ and $O_Q(0)$ instead of $\Phi_b^j$ where $O_P$, $O_Q$ are operators defined on some regions $P$, $Q$ we get
\begin{flalign}
\norm{\com{O_P(t)}{O_Q(0)}}  \leq N \norm{O_P}\norm{{O_Q}} M \times \sum_{n=0}^\infty \frac{(2 \abs{t} \sqrt{h_{a} h_{a+1}})^n}{n!}\underbrace{\Big(\sum_{i_{2}\in Z_{i_1}}\ldots\sum_{i_{n+1}\in Z_{i_n}}\delta^{i_n}_Q\Big)}_{c_n} , \label{binductivetoo}
\end{flalign}
where $M$ is defined as previously described and $N$ is the (finite) number of local operators of the Hamiltonian intersecting $P$. Here $c_n$ counts the number of chains associated to $n$ local operators (and thus $n$ subgraphs $\Gamma(c,k)$), starting with $\Gamma(a,i_1)$, which must intersect $P$, and ending with $\Gamma(b,i_n)$, which must intersect $Q$.  Furthermore, we can always find a bound of the following type for $c_n$:
\begin{align}
c_n \leq \tilde{M} \gamma^n e^{\lambda(\frac{n}{\xi} - d)} \label{cng1},
\end{align}
where $\lambda$ is an arbitrary positive real number.
This is because the $\Gamma(a,i)$'s have a diameter of $R$ or less. Hence, if the distance $d$ between the initial and final point is greater than $nR$ then there are no possible chains of $n$ local operators linking the initial and final point. Since there are always fewer than $\nu$ other local operators intersecting $\Gamma(a,i)$  ($\nu$ must be finite because all vertices have a valence of not more than $\zeta$), there is a maximum of $\nu^n$ possible chains of $n$ local operators starting from any given position.  Thus, we certainly have that
\begin{align}
c_n \leq  \nu^n e^{\lambda(R {n} - d)} \label{cngex}.
\end{align}
Using \rf{cng1} in \rf{binductivetoo} we obtain
\begin{align}
\begin{split}
\norm{\com{O_P(t)}{O_Q(0)}}  \leq  \tilde{\tilde{M}} &\exp{\lambda\left(2\sqrt{h_0 h_1}\frac{\gamma}{\lambda} e^{\frac{\lambda}{\xi}}t-  d\right)},\end{split} \label{Kzobiwan}
\end{align}
where $\tilde{\tilde{M}}= M \tilde{M} N \norm{O_P}\norm{{O_Q}}$. This gives us an upper bound on the speed information can travel
\begin{align}
v_{LR}= 2\sqrt{h_0 h_1}\frac{\gamma}{\lambda} e^{\frac{\lambda}{\xi}},
\end{align}
which can be minimised by choosing $\lambda=\xi$. We thus get the following upper bound on the speed information can travel
\begin{align}
v_{LR}= 2\frac{\gamma}{\xi} e\sqrt{h_0 h_1} .\label{vlrgenobiwan}
\end{align}

\subsection{Lieb-Robinson Bound on a Lattice} \label{sec_LR_Lattice}

In the case where the graph is a regular lattice, it is possible to give a bound of the form given in \rf{cng1} as a function of the lattice and the Hamiltonian.  We show here how this can be done.

For the Hamiltonian given in \rf{deuxint} on a regular lattice we have the following properties (in addition to the properties on a general graph):  By passing from a  $\Phi_0$ to a $\Phi_1$ in an operator chain, one always moves along the lattice by a lattice distance of $D_1$ and by passing from a  $\Phi_1$ to a $\Phi_0$ in an operator chain, one invariably moves along the lattice by a lattice distance of $D_0$; in addition, for a given $i$, the finite number $n_{a\rightarrow b}$ of $j$'s such that $\Gamma(a,i)$ and $\Gamma(b,j)$ have a non-empty intersection is independent of $i$ and $j$ and depends only on $a$ and $b$, that is, the interaction types.
Since the chains must alternate between operators of type $0$ and operators of type $1$, the number of operator chains of length $n$ starting at a specified point on the lattice, but with an unconstrained end point, is, at most,
\begin{align}
c_n\leq \max\left\{\sqrt{\frac{n_{0\rightarrow 1}}{n_{1\rightarrow 0}}}, \sqrt{\frac{n_{1\rightarrow 0}}{n_{0\rightarrow 1}}}\right\} \sqrt{n_{0\rightarrow 1} n_{1\rightarrow 0}}^n.\label{blanc}
\end{align}
where the max prefactor takes care of both even and odd chains. In this case, for $d(P,Q)>\frac{n+1}{2}(D_0 +D_1)$,
there cannot be any chains of $n$ operators linking the initial and final point. This means that for $\lambda>0$, $e^{\lambda \big(\frac{n+1}{2}(D_0 +D_1)- d(P,Q)\big)}$ is necessarily greater than $1$ if there exists at least one operator chain of length $n$ linking $P$ and $Q$. Thus, instead of \rf{blanc} we can write
\begin{align}
c_n\leq \tilde{M} \sqrt{n_{0\rightarrow 1} n_{1\rightarrow 0}}^n e^{\lambda \big(\frac{n}{2}(D_0 +D_1)- d(P,Q)\big)},\label{orpheline}
\end{align}
where $\tilde{M}= \max\left\{\sqrt{\frac{n_{0\rightarrow 1}}{n_{1\rightarrow 0}}}, \sqrt{\frac{n_{1\rightarrow 0}}{n_{0\rightarrow 1}}}\right\}e^{\frac{\lambda}{2}(D_0 +D_1)}$. Referring back to \rf{cng1} we obtain:
\begin{align}
\gamma = \sqrt{n_{0\rightarrow 1} n_{1\rightarrow 0}}, \label{gammagen}\\
\xi = \frac{2}{D_0 + D_1} .\label{xigen}
\end{align}
Combining these results with \rf{vlrgenobiwan} gives
\begin{align}
v_{LR}=  e\sqrt{n_{0\rightarrow 1} n_{1\rightarrow 0} h_0 h_1}(D_0 + D_1). \label{vlattice}
\end{align}

Of course, the restriction to lattices is a significant one; however, as we see in what follows, it is possible to obtain an equivalent bound on $c_n$ for more general graphs.

\subsection{Lieb-Robinson Bound on a Homogeneous and Isotropic Graph} \label{sec_LR_HOMO}

Let us now suppose that we have a general graph, and not necessarily a lattice. In this case, the $D$'s of Section \ref{sec_LR_Lattice} gain a dependence on the specific local operators $\Phi^i$ and $\Phi^j$; that is, instead of simply $D_0$ and $D_1$, we have $D_0^{i\rightarrow j}$ and $D_1^{j\rightarrow k}$. Similarly, the $n$'s gain a dependence on $i$: Instead of simply $n_{0\rightarrow 1}$ and $n_{1\rightarrow 0}$, we have $n_{0\rightarrow 1}^i$ and $n_{1\rightarrow 0}^j$.  For a given operator chain of $n$ operators $\{\phi_a^{i_1}, \phi_{a+1}^{i_2}, \ldots , \phi_b^{i_{n-1}},\phi_{b+1}^{i_n}\}$, the end points of the chain can be separated by, at most, a graph distance of $\sum_{k=1}^{n-1}D_{a_k}^{i_k\rightarrow i_{k+1}}$. Hence, if $d(P,Q)> \sum_{k=1}^{n-1}D_{a_k}^{i_k\rightarrow i_{k+1}}$, the chain cannot link $P$ to $Q$. 

Let us consider chains of $n$ operators starting at $i_1$. Let $i_k$ and $j_k$ be different chains such that $i_1=j_1$.  Then if we have
\begin{align}
 \frac{1}{n}\sum_{\mbox{chain i }k=1}^{n-1}D_{a_k}^{i_k\rightarrow i_{k+1}} & \approx \frac{1}{n}\sum_{\mbox{chain j }k=1}^{n-1}D_{a_k}^{j_k\rightarrow j_{k+1}} \mbox{ and}, \label{homogenous} \\
 \sqrt[n]{\prod_{\mbox{chain i }k=1}^{n-1}n_{a_k\rightarrow a_k+1}^{i_k}} & \approx \sqrt[n]{\prod_{\mbox{chain j }k=1}^{n-1}n_{a_k\rightarrow a_k+1}^{j_k}} ,\label{isotropic}
 \end{align}
and so the Lieb-Robinson speed will be the same in all directions.  By $\approx$ we mean that the quantities converge for large $n$ and are thus independent of the path the chain makes on the lattice.  Moreover, if the quantities in Eqs. (\ref{homogenous}) and (\ref{isotropic}) depend on $i_n$, then the Lieb-Robinson speed will vary in different regions of the graph. Finally, if the quantities in Eqs. (\ref{homogenous}) and (\ref{isotropic}) depend on $n$, then the speed will depend on the distance $d$.

We say a graph is homogeneous if the quantities in Eqs. (\ref{homogenous}) and (\ref{isotropic}) do not depend on $n$.  Furthermore, we say a graph is isotropic at $i_1$ if Eqs. (\ref{homogenous}) and (\ref{isotropic}) do not depend on $i_n$, in which case the Lieb-Robinson speed will be the same in all directions. If these conditions are satisfied, the Lieb-Robinson speed will be constant in all directions and areas of the graph (or subgraph) considered.

It follows that in the case of a homogeneous and isotropic graph,  we can define
 \begin{align}
 \nnn = \sqrt[n]{\prod_{\mbox{chain i }k=1}^{n-1}n_{a_k\rightarrow a_k+1}^{i_k}}, \\
 \DDD = \frac{1}{n}\sum_{\mbox{chain i }k=1}^{n-1}D_{a_k}^{i_k\rightarrow i_{k+1}} ,\label{nbar}
 \end{align}
for some chain of length $n$ where $n$ is large.  By definition, these quantities will not depend on the particular chain chosen.  We then have that $\nnn^n$ is equal to the number of chains of $n$ operators starting at a given point and ending anywhere. Note that for $d(P,Q)>n\DDD$
there cannot be any chains of $n$ operators linking the initial and final points. This means that for $\lambda>0$, $e^{\lambda \big(n\DDD- d(P,Q)\big)}$ is necessarily greater than $1$ if there exists at least one operator chain of length $n$ linking $P$ and $Q$. Thus, we can write the analogue of \rf{orpheline} by
\begin{align}
c_n\leq  \nnn^n e^{\Big(\lambda \big(n\DDD- d(P,Q)\big)\Big)},\label{ouchigea}
\end{align}
which gives
\begin{align}
\gamma = \nnn, \label{gammahomo}\\
\xi = \frac{1}{\DDD} .\label{xihomo}
\end{align}
Combining these results with \rf{vlrgenobiwan} gives us
\begin{align}
v_{LR}=  2 e\DDD\nnn\sqrt{ h_0 h_1}. \label{vhomo}
\end{align}
As in Sections \ref{sec_LR_Lattice}, we require a bound for the coefficient $c_n$ in order to provide a Lieb-Robinson speed in terms of the parameters of the system. Let
\begin{align}
 \nnn  =  \begin{cases}
\sqrt{n_{0\rightarrow 1} n_{1\rightarrow 0} },& \text{lattice},  \\
\sqrt[n]{\prod_{\mbox{chain i }k=1}^{n-1}n_{a_k\rightarrow a_k+1}^{i_k}}, & \text{h\&i graph},   \end{cases} \nn
 \DDD  =  \begin{cases}
\frac{1}{2}(D_0 +D_1), & \text{lattice},  \\
\frac{1}{n}\sum_{\mbox{chain i }k=1}^{n}D_{a_k}^{i_k\rightarrow i_{k+1}}, & \text{h\&i graph},  \end{cases} \label{cond}
\end{align}
where ``h\&i" stands for homogeneous and isotropic. Using the definitions of \rf{cond}, we have that \rf{vhomo} is the general equation for the Lieb-Robinson speed for both lattice systems and homogeneous and isotropic systems.
\subsection{Example: The Ising Model}

The quantum Ising model describes a spin network with a preferred alignment that is subjected to a magnetic field transverse to this preferred direction.  In one dimension, the model is exactly solvable and is known to exhibit a second-order phase transition at zero temperature \cite{Sachdev}.  The quantum Ising chain has a wide range of applications from condensed-matter physics \cite{Stinchcombe} to quantum gravity \cite{Berman}.

The quantum Ising chain is an example of a bounded system on a lattice having a Hamiltonian with two coupling constants.  Accordingly, we can use the framework developed previously to calculate the Lieb-Robinson bound.

The Hamiltonian for the quantum Ising chain is given by
\begin{equation}
  H = -J \sum_i \left( g \sigma_{i}^x + \sigma_{i}^{z} \sigma_{i+1}^{z} \right),
\end{equation}
where $\sigma^x$ and $\sigma^z$ are Pauli matrices.  At zero temperature the Ising chain exhibits two phases: paramagnetic ($g>1$) and ferromagnetic ($g<1$) with a quantum critical point at $g=1$.  By varying $g$, that is, the transverse magnetic field, the system can undergo a quantum phase transition from the paramagnetic phase (disordered) to the ferromagnetic phase (ordered) or vice versa.

In order to calculate the Lieb-Robinson speed we need an upper bound on the coefficients $c_n$ defined in \rf{cng1}.  The $c_n$ coefficients represent the number of paths of length $n$ consisting of alternating $\sigma_i^x$ and $\sigma_{i}^z \sigma_{i+1}^z$ operators such that the subgraphs corresponding to the operators intersect (in this case only nearest neighbours).  Since we are dealing with a lattice we can use \rf{gammagen} and \rf{xigen}. There are 2 possible steps from a vertex to an edge and 2 possible steps from an edge to a vertex. Thus by \rf{gammagen}, in this case, $\gamma= \sqrt{2\times 2} = 2$.  Furthermore, by passing from a vertex to an edge, and vice versa, we move by half of a lattice spacing, thus by virtue of \rf{xigen}, $\xi=2$.

It follows from Eq. (\ref{vlrgenobiwan}) that the Lieb-Robinson bound for the speed of propagation in the quantum Ising model is found to be

\begin{align}
v_{LR}= 2 e \sqrt{ g } J.
\end{align}

Close to the quantum phase transition (which happens at $g=1$), at small deviations from criticality, $|g-1|\ll 1$, it is possible to take the continuum limit which gives a relativistic Majorana fermion field theory in one dimension with a speed of light of $c=J$ and a mass of $m=(g-1)J$ \cite{gogol}. This means at up to a constant of order 1, $2 e$, which we expect due to the approximations made, the Lieb-Robinson gives the right value for the speed of information propagation, at least near the phase transition where it can be checked.

\section{Theories of Finitely Many Interactions} \label{sec_three_interactions}

Now that we have established in detail how to calculate the Lieb-Robinson bound for two interactions, we will apply the intuition we gained to derive a bound for any finite number of interactions. The essential intuition which we gained for two interactions was that, in an operator chain, because one type of interaction commutes with interactions of the same type, each operator in a chain must be of a different type than the preceding one.  In the case of only two types of interaction, this requirement uniquely defines the interaction type of each operator in the chain; in other words, they must alternate. However, in the case of $m$ interactions, the situation is more complicated because we have a choice of as many as $m-1$ interaction types for each operator along the chain. The way we will deal with this added complication is by considering periodic chains of operators. Since the bound for an arbitrary chain will be the geometric mean of the bounds for such periodic chains, a bound on any chain can be derived from the bounds on the periodic chains.

Let us now consider general Hamiltonians possessing $m$ types of interactions.  We define a type of interaction to be a subset of the the local operators making up the Hamiltonian such that all the operators commute with each other and such that they appear with the same coupling constant in the Hamiltonian. Intuitively, we think of all the operators of a single type as performing the same action at different locations on the graph.  For example, in the quantum Ising model, the set of local operators $\{\sigma^x_i\}_{i\in \Z}$ implementing the transverse magnetic field is considered to be one type of interaction. We will derive the Lieb-Robinson bound for such a Hamiltonian and provide the specific dependence of the Lieb-Robinson speed on the strength of the interactions. For simplicity, we will restrict ourselves to the case of bounded operators on a lattice.  Generalisations to the commutator-bounded case and to homogeneous and isotropic graphs are straightforward applications of the methods shown in the previous sections, the results for which we state at the end of this section.

We consider a Hamiltonian of the following type:
\begin{align}
H \equiv \sum_{l=1}^m\sum_{i_l\in S_l} h_l \Phi_l^{i_l},
\end{align}
where $l$ labels the interaction type.   We assume that all the terms in the sum are non-zero and that for all $l\in \{1,\ldots,m\}$ and $\forall (i,j)\in S_l^2$ we have $\com{\Phi_l^i}{\Phi_l^j}=0$ and $\norm{\Phi_l^i}=\norm{\Phi_l^j} = 1$.  Note that the normalization of the operators can be imposed without any loss of generality in the case of bounded interactions.  Further, we assume that for all  $(l_1,l_2)\in \{1,\ldots,m\}^2$ and $i\in S_{l_1}$, there are exactly $n_{l_1\rightarrow l_2}$ $j$'s such that $j\in S_{l_2}$ and $\Gamma(l_1,i)\cap\Gamma(l_2,j) \neq \emptyset$, where $n_{l_1\rightarrow l_2}$ depends only on $l_1$ and $l_2$. In addition, the graph distance travelled along the graph by passing from $\Gamma(l_1,i)$ to one such $\Gamma(l_2,j)$ is $D_{l_1}^{l_1\rightarrow l_2}$, which also depends only on $l_1$ and $l_2$. The last two properties arise from the fact that we are supposing the graph to be a lattice.

\subsection{Lieb-Robinson Bound}

For Hamiltonians with two interaction types it was easy to calculate the Lieb-Robinson bound.  This is because in order for a signal to propagate the two types of interactions had to alternate in the operator chains which showed up in the bound. Therefore, we knew that half the operators were of one type of interaction while the other half were of the other type. For $m$ types of interactions, different chains could have many different combinations of each type of operator, the only requirement being that each operator in the chain implements a different interaction than the previous operator. The intuition is then to consider all possible chains which are periodic in the interaction type and then bound all chains by the periodic sequence of operators which has the maximal norm. This will work because even if a chain is not periodic, it can be subdivided into components which are subchains of periodic chains.

Everything up to \rf{binductive} does not depend on the number of interaction types. In the case of $m$ interactions, \rf{binductive} becomes
\begin{align}
 \norm{{K_{a \ b}^{i_1 \ j}(t)}}  \leq M \sum_{n=0}^\infty \frac{\abs{t}^n}{n!}\Big(\sum_{i_{2}\in Z_{i_1}}2 h(i_2)\ldots\sum_{i_{n+1}\in Z_{i_n}}2 h(i_{n+1}) \delta^{i_n}_j\Big) , \label{trinductive}
\end{align}
where $h(i_k)\in\{h_1, h_2, \ldots, h_m\}$ and of course $h(i_k) \neq h(i_{k+1})$. This situation is now more complicated than for two interactions because there is not necessarily the same number of each coupling constant in each term. The power of a particular coupling constant in a particular term depends on the operator chain and can range anywhere from $0$ to $\lceil{n/2}\rceil$.

In \rf{trinductive} the end points of the chain are separated by, at most, a graph distance of $\sum_{k=1}^{n-1}D_{a(i_k)}^{a(i_k)\rightarrow a(i_{k+1})}$, where $a(i_k)\in\{1,2,\ldots,m\}$ labels the operator type of $i_k$. Hence, if $d(P,Q)> \sum_{k=1}^{n-1}D_{a(i_k)}^{a(i_k)\rightarrow a(i_{k+1})}$, the chain cannot link $P$ to $Q$. As we did previously, we can bound $\norm{{K_{a \ b}^{i_1 \ j}(t)}}$ by multiplying every term in \rf{trinductive} by $\exp\Big(\lambda \big(\sum_{k=1}^{n-1}D_{a(i_k)}^{a(i_k)\rightarrow a(i_{k+1})}- d(P,Q)\big)\Big)$, which takes care of the end-point condition (i.e., the $\delta^{i_n}_j$ factor in \rf{trinductive}). By doing this, it can be shown (see Appendix A for details) that the Lieb-Robinson bound takes the form
\begin{align}
 \norm{\com{O_P(t)}{O_Q(0)}}   \leq  \tilde{M}(\lambda)\exp \Big(\lambda \big( \frac{2(m-1)L(\lambda)}{\lambda}\abs{t} - d(P,Q)\big) \Big),\label{LRB3}
\end{align}
where $\lambda$ is again an arbitrary positive number, $\tilde{M}<\infty$ for finite $\lambda$ and
\begin{align}
 L(\lambda) \equiv \max_{2\leq r\leq m(m-1)} \max_{(a_1, a_2, \ldots, a_r)\in \{1,\ldots,m\}^r} \bigg\{k_{a_1 \ldots a_r}e^{\lambda \xi_{a_1 \ldots a_r}}|  1\leq p<q\leq r, (a_p, a_{p+1})\neq (a_q,a_{q+1})\bigg\} , \label{Ltext}
\end{align}
where
\begin{align}
k_{a_1 a_2 \ldots a_q} & \equiv \sqrt[q]{\prod_{l=1}^q h_{a_l}n_{a_l\rightarrow a_{l+1}}}, \nn
\xi_{a_1 a_2 \ldots a_q} & \equiv  \frac{\sum_{l=1}^q D_{a_{l}}^{a_l\rightarrow a_{{l+1}}}}{q},\label{minicycltext}
\end{align}
and where we assume $a_{r+1}\equiv a_1$.  This implies a Lieb-Robinson speed of
\begin{align}
 v_{L R} =  \inf_{\lambda>0} \frac{2(m-1)L(\lambda)}{\lambda}.\label{vlrtext}
\end{align}

We can generalise this result to homogeneous and isotropic graphs using the same methods used in the previous sections.  All that is needed is to redefine the $k$'s and $\xi$'s in \rf{minicycltext}. By the definition of a homogeneous and isotropic graph, the result of taking the arithmetic mean of the $D$'s and the geometric mean of the $n$'s over a chain will not depend on the chain taken. In other words,
if we let $C_o(P,x,l_1,l_2,\ldots,l_r)$ be the set of operator chains of $x$ operators whose first operator's support intersects $P$ and is of type $l_1$ and whose $q^{th}$ operator is of type $l_g$ where $g=[(q -1) \mod r]+1$, we may define:
\begin{align}
\nnn_{l_1 l_2 \ldots l_r} & = \sqrt[|C_o(P,x,l_1,l_2,\ldots,l_r)|]{\prod_{i\in C_o(P,x,l_1,l_2,\ldots,l_r)} \prod_{k= 1 }^{r}n_{l_k\rightarrow l_{k+1}}^{i_{f r+ k}}} ,\nn
\DDD_{l_1 l_2 \ldots l_r} & = \frac{1}{|C_o(P,x,l_1,l_2,\ldots,l_r)|}{\sum_{i\in C_o(P,x,l_1,l_2,\ldots,l_r)}\sum_{k= 1 }^{r}D_{l_k\rightarrow l_{k+1}}^{i_{f r+ k}\rightarrow i_{f r+ k+1}}}. \label{threesome}
\end{align}
We now summarise our results for the Lieb-Robinson bound for a system of $m$ interactions on a lattice or a homogeneous and isotropic (h\&i) graph.  The derivation was for bounded operators but we also state the result for commutator bounded systems which are discussed in \cite{LRB2}.  To state all results concisely we define a constant $C$ and state the forms of $k_{l_1 \ldots l_r}$ and $\xi_{l_1 \ldots l_r}$ for the different possible systems:
\begin{align}
 C  & = \begin{cases}2(m-1) & \text{bounded,} \\
                  \sqrt{8} (m-1) & \text{commutator-bounded,}
                  \end{cases} \label{mauditecste}\\
 k_{l_1 \ldots l_r} & = \begin{cases}\sqrt[r]{\prod_{i=1}^r h_{l_i} n_{l_i\rightarrow l_{i+1}}} & \text{bounded, lattice,}\\
                        \sqrt[r]{\prod_{i=1}^r h_{l_i}}\nnn_{l_1 l_2 \ldots l_r}  & \text{bounded, h\&i graph,}\\
                        \sqrt[r]{\prod_{i=1}^r h_{l_i}
                        K_{l_{i} l_{i+1}} n_{l_i\rightarrow l_{i+1}}} & \text{commutator-bounded, lattice,}\\
                        \sqrt[r]{\prod_{i=1}^r h_{l_i}} \nnn_{l_1 l_2 \ldots l_r} & \text{commutator-bounded, h\&i graph,}
\end{cases}\label{mauditk}\\
 \xi_{l_1 \ldots l_r} &  = \begin{cases}\frac{\sum_{i=1}^r D^{l_{i}\rightarrow l_{i+1}}}{r} & \text{lattice,}\\
                       \DDD_{l_1 \ldots l_r}  & \text{h\&i graph.}
                       \end{cases} \label{mauditxi}
\end{align}
For commutator bounded systems we define $K_{l_{i} l_{i+1}} = \|[\Phi_{l_i}, \Phi_{l_{i+1}}]\|$ for $(l_i,l_{i+1}) \in \{1,\ldots,m\}^2$.  With these redefinitions, the general version of the Lieb-Robinson bound and speed are
\begin{align}
 \norm{\com{O_P(t)}{O_Q(0)}}   \leq  \tilde{M}(\lambda)\exp \Big(\lambda \big( \frac{C L(\lambda)}{\lambda}\abs{t} - d(P,Q)\big) \Big),\label{LRB3g}
\end{align}
giving that
\begin{align}
 v_{L R} =  \inf_{\lambda>0} \frac{C L(\lambda)}{\lambda}.\label{vlrtextg}
\end{align}

\subsection{Calculating the Lieb-Robinson Speed}

Using Eqs. (\ref{mauditk}), (\ref{mauditxi}), and (\ref{vlrtextg}) we will calculate the Lieb-Robinson speed for a system of $m$ interactions.  If there are only two interactions labeled as 0 and 1, then the max is over a set containing only one element so we have
\begin{align}
v_{L R} = C \inf_{\lambda>0}\left\{\frac{k_{01}}{\lambda} e^{\xi_{01}\lambda} \right\}= C e k_{01} \xi_{01},
\end{align}
which agrees with our previous result for systems with two interactions. For more than two interactions the max is taken over a set containing more than one element which brings about a qualitative difference: One obtains a different functional dependence on the strengths of the interactions (the $h$'s), as well as the structure of the graph (the $n$'s and $D$'s) as compared to the algebraic dependence in the case of two interactions.

To calculate $v_{L R}$ for more than two interactions, we must take the maximum of a set of functions of the form $k_{i_1 \ldots i_r} e^{\xi_{i_1 \ldots i_r}\lambda}$.  Since the maximal element of this set can change for different values of $\lambda$, some care must be taken. In view of \rf{vlrtextg} we divide these elements by $\lambda$ since it does not change which element is maximal and we call this set $A$.  That is

\begin{align}
  A = \left\{ \frac{k_{i_1 \ldots i_r} e^{\xi_{i_1 \ldots i_r}\lambda}}{\lambda} \right\},
\end{align}
for the different possible subchain labellings $i_1 \ldots i_r$, as given in \rf{Ltext}.  Further, we define a function $A_{\max}(\lambda) = \max A$.  Therefore

\begin{align}
 v_{L R} =  C \inf_{\lambda>0} A_{\max}(\lambda).
\end{align}
The element of $A$ which has the largest value of $k$ will be maximal for $\lambda \rightarrow 0$.  The value of $\lambda$ at which two elements of $A$ intersect is given by $\frac{k_{l_1\ldots l_r}e^{\xi_{l_1\ldots l_r}\lambda}}{\lambda} = \frac{k_{j_1\ldots j_s}e^{\xi_{j_1\ldots j_s}\lambda}}{\lambda}$ and corresponds to the value

\begin{align}
\lambda^{{}_{(l_1\ldots l_t)}}_{{}_{(j_1\ldots j_s)}} = \frac{\ln(k_{l_1\ldots l_t})-\ln(k_{j_1\ldots j_s})}{\xi_{j_1\ldots j_s}- \xi_{l_1\ldots l_t}}. \label{lambada}
\end{align}
Notice that for $\lambda^{{}_{(l_1\ldots l_t)}}_{{}_{(j_1\ldots j_s)}}$ to be positive, we must have $\xi_{j_1\ldots j_s} > \xi_{l_1\ldots l_t}$, which is also the condition for the two curves to intersect.  Now suppose that $k_{i_1 \ldots i_r}$ is the largest value of $k$ of the elements in $A$ and that $\lambda^{(j_1\ldots j_s)}_{(i_1 \ldots i_r)}$ is the smallest intersection point involving this element.  Next, suppose that $\lambda^{{}_{(l_1\ldots l_t)}}_{{}_{(j_1\ldots j_s)}}$ is the smallest intersection point greater than $\lambda^{{}_{(j_1\ldots j_s)}}_{{}_{(i_1 \ldots i_r)}}$ involving the element labeled by $j_1 \ldots j_s$.  Continuing in this way, the function $A_{\max}(\lambda)$ will be given by

\begin{align}
  A_{\max}(\lambda) =
\begin{cases}
  \frac{k_{i_1 \ldots i_r}e^{\xi_{i_1 \ldots i_r}\lambda}}{\lambda}, & \hspace{29pt} \text{ $0 < \lambda \leq \lambda^{{}_{(j_1\ldots j_s)}}_{{}_{(i_1 \ldots i_r)}}$},  \\
  \frac{k_{j_1 \ldots j_s}e^{\xi_{j_1 \ldots j_s}\lambda}}{\lambda}, & \text{ $\lambda^{{}_{(j_1\ldots j_s)}}_{{}_{(i_1 \ldots i_r)}} \leq \lambda \leq \lambda^{{}_{(l_1\ldots l_t)}}_{{}_{(j_1 \ldots j_s)}}$}, \\
  \hspace{30pt} \vdots
\end{cases}
\end{align}
Since the function $\frac{k_{j_1 \ldots j_s}e^{\xi_{j_1 \ldots j_s}\lambda}}{\lambda}$ has only one minimum at $\lambda = 1/\xi_{j_1 \ldots j_s}$, it follows that if $\lambda^{{}_{(j_1\ldots j_s)}}_{{}_{(i_1 \ldots i_r)}} \leq 1/\xi_{j_1 \ldots j_s} \leq \lambda^{{}_{(l_1\ldots l_t)}}_{{}_{(j_1 \ldots j_s)}}$ for any one of the intervals, then the infimum of $A_{\max}(\lambda)$ is given by $A(1/\xi_{j_1 \ldots j_s})$.  Otherwise, the infimum of $A_{\max}(\lambda)$ will be at one of the points $\lambda^{{}_{(l_1\ldots l_t)}}_{{}_{(j_1\ldots j_s)}}$.  This completes the algorithm for finding the Lieb-Robinson speed $v_{L R}$ for any finite number of interactions.

For concreteness we work out some general scenarios for the case of three interactions.  We label the interactions by 1,2, and 3 in which case

\begin{align}
  A &= \Bigg\{ \frac{k_{1 2}e^{\xi_{1 2}\lambda}}{\lambda}, \frac{k_{2 3}e^{\xi_{2 3}\lambda}}{\lambda}, \frac{k_{3 1}e^{\xi_{3 1}\lambda}}{\lambda}, \frac{k_{1 2 3}e^{\xi_{1 2 3}\lambda}}{\lambda}, \frac{k_{1 3 2}e^{\xi_{1 3 2}\lambda}}{\lambda}, \nonumber \\
  & \hspace{20pt} \frac{k_{2 1 3 1}e^{\xi_{2 1 3 1}\lambda}}{\lambda}, \frac{k_{1 2 3 2}e^{\xi_{1 2 3 2}\lambda}}{\lambda}, \frac{k_{1 3 2 3}e^{\xi_{1 3 2 3}\lambda}}{\lambda} , \frac{k_{1 2 3 1 3 2}e^{\xi_{1 2 3 1 3 2}\lambda}}{\lambda} \Bigg\}.
\end{align}
Note that we have only considered distinct elements of the set.  For example, we did not consider $\frac{k_{2 3 1}e^{\xi_{2 3 1}\lambda}}{\lambda}$ because $\frac{k_{2 3 1}e^{\xi_{2 3 1}\lambda}}{\lambda} = \frac{k_{1 2 3}e^{\xi_{1 2 3}\lambda}}{\lambda}$.

There are two possible scenarios.  The first is that one of the $k$'s and its associated $\xi$ are greater than all the other $k$'s and $\xi$'s.  In that case, one element of $A$ will be greater than all the others regardless of the value of $\lambda$ and the Lieb-Robinson speed will be given by minimising that particular element of $A$.  If, for example $k_{1 3 2}$ is greater or equal to all the other $k$'s while simultaneously $\xi_{1 3 2}$ is greater or equal to all the other $\xi$'s, then $\frac{k_{1 3 2}e^{\xi_{1 3 2}\lambda}}{\lambda}$ will be the maximum element irrespective of $\lambda$, and so we have
\begin{align}
v_{L R} & = C \inf_{\lambda>0}\frac{k_{1 3 2}e^{\xi_{1 3 2}\lambda}}{\lambda}, \nonumber \\
& = Ce k_{1 3 2}\xi_{1 3 2}. \label{terrorist}
\end{align}
For a bounded system on a lattice $C = 2(m-1)$ and

\begin{align}
  k_{123} & = \sqrt[3]{h_1 h_2 h_3} \sqrt[3]{n_{1\rightarrow 3} n_{3 \rightarrow 2} n_{2 \rightarrow 1}}, \\
  \xi_{123} & = \frac{D^{1\rightarrow 3}+ D^{3\rightarrow 2}+ D^{2\rightarrow 1}}{3},
\end{align}
which allows us to solve for the Lieb-Robinson speed explicitly in terms of the coupling constants and lattice distances as

\begin{align}
v_{L R} & = \frac{4 e}{3}  \sqrt[3]{h_1 h_2 h_3}\sqrt[3]{n_{1\rightarrow 3} n_{3 \rightarrow 2} n_{2 \rightarrow 1}}(D^{1\rightarrow 3}+ D^{3\rightarrow 2}+ D^{2\rightarrow 1}).\nn
\end{align}

In the second scenario, no single element of $A$ is greater than all the others for all values of $\lambda$. Let us suppose, in that case, that we have $k_{1 2} > k_{2 3} > k_{1 2 3}> k_{3 1} > k_{1 3 2}>k_{1 2 3 2}>k_{1 2 3 1 3 2}>k_{2 1 3 1}>k_{1 3 2 3}$ and  $\xi_{1 2 3} > \xi_{2 3} > \xi_{1 2}> \xi_{1 3 2 3}>\xi_{2 1 3 1}>\xi_{1 2 3 1 3 2} > \xi_{1 2 3 2}>\xi_{3 1} > \xi_{1 3 2}$.  With these assumptions one finds that

\begin{align}
  A_{\max}(\lambda) =
\begin{cases}
  \frac{k_{12}e^{\xi_{12}\lambda}}{\lambda}, & \hspace{18pt} \text{ $0 < \lambda \leq \lambda^{{}_{(23)}}_{{}_{(12)}}$}, \\
  \frac{k_{23}e^{\xi_{23}\lambda}}{\lambda}, & \hspace{3pt} \text{ $\lambda^{{}_{(23)}}_{{}_{(12)}} \leq \lambda \leq \lambda^{{}_{(123)}}_{{}_{(23)}}$}, \\
  \frac{k_{123}e^{\xi_{123}\lambda}}{\lambda}, & \text{ $\lambda^{{}_{(123)}}_{{}_{(23)}} \leq \lambda \leq \ldots$}, \\
  \hspace{19pt} \vdots
\end{cases}
\end{align}
and also that $\xi_{12}^{-1}$, $\xi_{23}^{-1}$, and $\xi_{123}^{-1}$ do not lie in the intervals of $\lambda$ for which their corresponding element in $A$ is maximal, i.e. none of these are the minimal point of $A_{\max}(\lambda)$.

In Table \ref{c-chelou} we state the maximal element of $A$ in the different relevant intervals of $\lambda$.  We also indicate whether the function $A_{\max}(\lambda)$ is increasing or decreasing in each interval.  From the table we see that the function $A_{\max}(\lambda)$ is minimized for $\lambda = \lambda_{{}_{(2 3)}}^{{}_{(1 2 3)}}$.

\begin{table}
 \begin{tabular}{c|c|c|c|c|c|c|c|c|c|c|c|c|c|c|c}
\hline
 inc. or& \ & \multicolumn{2}{c|}{} & \multicolumn{2}{c|}{} & \multicolumn{2}{c|}{} & \multicolumn{2}{c|}{} & \multicolumn{2}{c|}{} & \multicolumn{2}{c|}{} & \multicolumn{2}{c}{}  \\
 dec.: & \  &\multicolumn{2}{c|}{$\searrow$} & \multicolumn{2}{c|}{$\searrow$} & \multicolumn{2}{c|}{$\ \ \searrow \ \ $} & \multicolumn{2}{c|}{$\ \ \searrow \ \ $} & \multicolumn{2}{c|}{$\nearrow$} & \multicolumn{2}{c|}{$\ \ \nearrow \ \ $}  & \multicolumn{2}{c}{$\nearrow$} \\
   &  \ & \multicolumn{2}{c|}{} & \multicolumn{2}{c|}{} & \multicolumn{2}{c|}{} & \multicolumn{2}{c|}{} & \multicolumn{2}{c|}{} & \multicolumn{2}{c|}{} & \multicolumn{2}{c}{}  \\
\hline
 max & \ & \multicolumn{1}{c}{$\frac{k_{1 2}e^{\xi_{1 2}\lambda}}{\lambda}$} & \ & \multicolumn{2}{c}{$\frac{k_{2 3}e^{\xi_{2 3}\lambda}}{\lambda}$} & \multicolumn{1}{c}{} & \multicolumn{1}{c}{} &\multicolumn{1}{c}{}  &\multicolumn{1}{c|}{}  &  \multicolumn{2}{c}{$\frac{k_{1 2 3}e^{\xi_{1 2 3}\lambda}}{\lambda}$} &\multicolumn{1}{c}{}&\multicolumn{1}{c}{}&\multicolumn{1}{c}{}&\multicolumn{1}{c}{} \\
 term: & \ & \multicolumn{2}{c|}{} & \multicolumn{2}{c}{} & \multicolumn{2}{c}{} & \multicolumn{2}{c|}{} & \multicolumn{2}{c}{} & \multicolumn{2}{c}{} & \multicolumn{2}{c}{}  \\\hline
 & \ & \multicolumn{2}{c|}{} & \multicolumn{2}{c|}{} & \multicolumn{2}{c|}{} & \multicolumn{2}{c|}{} & \multicolumn{2}{c|}{} & \multicolumn{2}{c|}{} & \multicolumn{2}{c}{}  \\
 $\lambda$ :&  \multicolumn{2}{l}{$0$} & \multicolumn{2}{l}{$\lambda_{{}_{(1 2)}}^{{}_{(2 3)}}$}  &  \multicolumn{2}{r}{$\ \xi_{123}^{-1}$}  & \multicolumn{2}{r}{$\ \lambda_{{}_{(1 2)}}^{{}_{(1 2 3)}}$}& \multicolumn{2}{r}{$\lambda_{{}_{(2 3)}}^{{}_{(1 2 3)}}$}& \multicolumn{2}{r}{$\xi_{2 3}^{-1}$}& \multicolumn{2}{r}{$\ \xi_{1 2}^{-1}$} &
\end{tabular}\caption{ The minimal point of the function $A_{\max}(\lambda)$ is located at $\lambda = \lambda_{{}_{(2 3)}}^{{}_{(1 2 3)}}$.  The top row indicates whether the function $A_{\max}(\lambda)$ is increasing or decreasing in the intervals of $\lambda$ determined by the various critical points.  The middle row indicates the maximal element of the set $A$ while the bottom row is a scale for $\lambda$ beginning at $\lambda=0$.  Also shown are $\xi_{12}^{-1}$, $\xi_{23}^{-1}$, and $\xi_{123}^{-1}$ which are the minima of their respective elements in $A$.  In this case they are not minima of $A_{\max}(\lambda)$ because they do not lie in the intervals for which their corresponding element of $A$ is maximal. } \label{c-chelou}
\end{table}
Thus we have
\begin{align}
v_{L R}  & = C \frac{k_{2 3}e^{\xi_{2 3}\lambda_{{}_{(2 3)}}^{{}_{(1 2 3)}}}}{\lambda_{{}_{(2 3)}}^{{}_{(1 2 3)}}}.
\end{align}
For a bounded system on a lattice we have $C = 2(m-1)$ and

\begin{align}
  k_{23} & = \sqrt{h_2 h_3} \sqrt{n_{2\rightarrow3}n_{3\rightarrow2}}, \\
  \xi_{23} & = \frac{D^{2\rightarrow 3} + D^{3\rightarrow 2}}{2},
\end{align}
which allows us to solve for the Lieb-Robinson speed explicitly in terms of the coupling constants and lattice distances as

\begin{align}
v_{L R} &  = 2 (m-1)\frac{(\xi_{1 2 3} - \xi_{2 3})}{\ln(k_{2 3})-\ln(k_{1 2 3})}  k_{2 3}^{\frac{\xi_{1 2 3}}{\xi_{1 2 3} - \xi_{2 3}}} k_{1 2 3}^{\frac{\xi_{2 3}}{\xi_{2 3} - \xi_{1 2 3}}}\label{alien}\\
&  = 4\frac{[\frac{{h_2 + h_3}}{2} - \frac{{h_1 + h_2 + h_3}}{3} ] + [\frac{n_{2 \rightarrow 3} + n_{3 \rightarrow 2}}{2} - \frac{n_{1\rightarrow 2}+ n_{2 \rightarrow 3}+ n_{3 \rightarrow 1}}{3}]}{\frac{D^{1\rightarrow 2}+ D^{2\rightarrow 3}+ D^{3\rightarrow 1}}{3}- \frac{D^{2\rightarrow 3}+ D^{3\rightarrow 2}}{2}} \nn
& \times \sqrt{h_2 h_3 n_{2 \rightarrow 3} n_{3 \rightarrow 2}}^{\frac{\frac{D^{1\rightarrow 2}+ D^{2\rightarrow 3}+ D^{3\rightarrow 1}}{3}}{\frac{D^{1\rightarrow 2}+ D^{2\rightarrow 3}+ D^{3\rightarrow 1}}{3}- \frac{D^{2\rightarrow 3}+ D^{3\rightarrow 2}}{2}}} \nn
&  \times \sqrt[3]{h_1 h_2 h_3 n_{1 \rightarrow 2} n_{2 \rightarrow 3} n_{3 \rightarrow 1}}^{\frac{\frac{ D^{2\rightarrow 3}+ D^{3\rightarrow 2}}{2}}{\frac{D^{2\rightarrow 3}+ D^{3\rightarrow 2}}{2} - \frac{D^{1\rightarrow 2}+ D^{2\rightarrow 3}+ D^{3\rightarrow 1}}{3}}} . \label{ahsijetait}
\end{align}

In general, when the minimum of $A_{\max}$ is attained at a minimum of one of the elements in $A$, the Lieb-Robinson speed will be of the form of \rf{terrorist}.  When the minimum of $A_{\max}$ occurs at the point where two elements of $A$ are equal, the Lieb-Robinson speed will be of the form of \rf{alien}.  Thus, the Lieb-Robinson speed will always be of one of the two forms
\begin{align}
v_{L R}^s(k,\xi) = 2 c e(m-1)k\xi \label{vlrs}
\end{align}
or
\begin{align}
 v_{L R}^d(k_1,k_2,\xi_1,\xi_2)= 2 c (m-1)\frac{(\xi_1 - \xi_2)k_1^{\frac{\xi_2}{\xi_2-\xi_1}}k_2^{\frac{\xi_1}{\xi_1-\xi_2}}}{\ln(k_2) - \ln(k_1)} , \label{vlrd}
\end{align}
where $c=1$ for bounded systems and $c=\sqrt{2}$ for commutator-bounded systems.  This is true also for any $m \geq 3$ since adding more interactions merely increases the number of elements of $A$.  Now that we have seen how to calculate the Lieb-Robinson speed for any number of interactions, let us consider a physical example.

\subsection{Example: XY-Model} \label{Section_XY}
We investigate the anisotropic quantum XY-model on a 2-dimensional square lattice.  This model is well understood and is studied mainly in connection with quantum phase transitions.  This model is also of practical interest, for example, in understanding the isolator-superconductor phase transition in cuprate type II superconductors \cite{supercond}.  At zero temperature the system exhibits two phases, one with an energy gap and another which is gapless.

The Hamiltonian for the quantum XY-model is given by
\begin{align}
H = J \sum_{\langle n,m\rangle} ( S_n^x S_m^x +  S_n^y S_m^y ) + D \sum_n (S_n^z)^2 ,\label{XYH}
\end{align}
where $\langle n,m \rangle$ represents the sum over nearest neighbours on the sites, $n$, of a square lattice.  Note that the gapless and gapped phases are determined by $\delta < \delta_c$ and $\delta > \delta_c$ respectively where $\delta= \frac{D}{2 J}$ and where $\delta_c$ has been estimated to be approximately 3.6 \cite{pires}.  For $\delta > \delta_c$, the correlation length of the system approaches a finite value leading to a quantum paramagnetic ground state with no long-range order.  We label the $S^x S^x$, $S^y S^y$, and $S^z$ operator types by X, Y, and Z respectively\footnote{Note that even though $S_n^x S_m^x$ and $S_n^y S_m^y$ have the same coupling constant, we must nevertheless consider them as two separate interactions because otherwise the conditions laid out between \rf{deuxint} and \rf{f} would not be satisfied.}.

Since the $k$'s and $\xi$'s are symmetric under cyclic permutations as well as the interchange of $X$ and $Y$, the set $A$ will have only four distinct elements:

\begin{align}
A = \left\{\frac{k_{X Y}e^{\lambda \xi_{X Y}}}{\lambda},\frac{k_{X Z}e^{\lambda \xi_{X Z}}}{\lambda}, \frac{k_{X Y Z}e^{\lambda \xi_{X Y Z}}}{\lambda},  \frac{k_{X Y Z Y}e^{\lambda \xi_{X Y Z Y}}}{\lambda} \right\}.
\end{align}
Note that $X Z Y Z = X Z$ and $X Y Z X Z Y = X Y Z$ by breaking the larger chains in two and using the symmetries.

The operators $X$ and $Y$ are located on edges of the graph, whereas the operators of type $Z$ are located on the plaquettes, the elementary squares of the lattice composed of four edges.  Moving from a plaquette to an edge corresponds to moving a lattice distance of 1/2, while moving from a plaquette to another plaquette corresponds to moving a lattice distance of 1.  Therefore

\begin{align}
 D_{X\rightarrow Z} &= D_{Y\rightarrow Z} = 1/2, \\
 D_{X\rightarrow Y} &= 1.
\end{align}
Note that these distances are symmetric.  Furthermore, from an edge one can move to one of two plaquettes or one of 7 other edges (including the one currently on).  From a plaquette one can move to one of four edges.  Therefore,

\begin{align}
 n_{X \rightarrow Z} &= n_{Y \rightarrow Z} = 2, \\
 n_{Z \rightarrow X} &= n_{Z \rightarrow Y} = 4, \\
 n_{X \rightarrow Y} &= n_{Y \rightarrow X} = 7.
\end{align}
Using this information and the coupling constants $h_X = J$, $h_Y = J$, and $h_Z = D$ in Eqs. (\ref{mauditk}) and (\ref{mauditxi}), we get

\begin{align}
  k_{X Y} & = 7 J S,& \xi_{XY} &= 1, \nn
  k_{X Z} & = \sqrt{8 JD}S,& \xi_{X Z} &= \frac{1}{2}, \nn
  k_{X Y Z} & =  \sqrt[3]{(7 \cdot 8) J^2 D} S,& \xi_{X Y Z} &= \frac{2}{3}, \nn
  k_{X Y Z Y} & = \sqrt[4]{(7^2 \cdot 8) J^3 D} S,& \xi_{X Y Z Y} &= \frac{3}{4},
\end{align}
where $S$ is a half integer giving us the type of spins on the lattice.  From \rf{lambada}, we obtain
\begin{align}
\lambda_{{}_{(X Z)}}^{{}_{(X Y)}} = \lambda_{{}_{(X Y Z)}}^{{}_{(X Y)}} = \lambda_{{}_{(X Y Z Y)}}^{{}_{(X Y)}} = \lambda_{{}_{(X Z)}}^{{}_{(X Y Z)}} = \lambda_{{}_{(X Z)}}^{{}_{(X Y Z Y)}} = \lambda_{{}_{(X Y Z)}}^{{}_{(X Y Z Y)}} =  2\ln(4/7) +\ln\left(\frac{D}{2 J}\right).\label{limbo}
\end{align}
This implies that all of the elements of $A$ intersect at the same value of $\lambda$.  First note that this value is negative for $D/2J < (7/4)^2$ and thus the curves do not intersect.  In this case, $\frac{k_{X Y}e^{\lambda \xi_{X Y}}}{\lambda}$ is maximal for all $\lambda$ and so the Lieb-Robinson speed is

\begin{align}
  v_{LR} &= 2e(m-1)k_{XY} \xi_{XY}, \nonumber \\
         &= 28eJS.
\end{align}
For $D/2J > (7/4)^2$ we have

\begin{align}
  A_{\max}(\lambda) =
\begin{cases}
  \frac{k_{XY}e^{\xi_{XY}\lambda}}{\lambda}, & \hspace{18pt} \text{ $0 < \lambda \leq \lambda^{{}_{(XZ)}}_{{}_{(XY)}}$},  \\
  \frac{k_{XZ}e^{\xi_{XZ}\lambda}}{\lambda}, & \text{ $\lambda^{{}_{(XZ)}}_{{}_{(XY)}} \leq \lambda \leq \infty$}. \\
\end{cases}
\end{align}
First let us consider the cases when the minima occur within the intervals.  The minimum occurs in the first interval when $0 < \xi_{XY}^{-1} \leq \lambda^{{}_{(XZ)}}_{{}_{(XY)}}$, which corresponds to $(7/4)^2 < D/2J < e(7/4)^2$ and gives the same speed as for $D/2J < (7/4)^2$.  The second interval contains the minimum when $\lambda^{{}_{(XZ)}}_{{}_{(XY)}} < \xi_{XZ}^{-1} \leq \infty$, which occurs for $D/2J > e^2(7/4)^2$.  In this case the minimum value is $A_{\max}(\xi_{XZ}^{-1})$ and gives a Lieb-Robinson speed of

\begin{align}
  v_{LR} &= 2e(m-1)k_{XZ} \xi_{XZ} \nonumber \\
         &= 4e\sqrt{2 JD}S.
\end{align}
Finally, in the region $e(7/4)^2 < D/2J < e^2(7/4)^2$ the minimum occurs for $\lambda = \lambda^{{}_{(XZ)}}_{{}_{(XY)}}$ and so

\begin{align}
  v_{LR} &= 2(m-1)\frac{k_{XY} e^{\xi_{XY} \lambda^{{}_{(XZ)}}_{{}_{(XY)}}}}{\lambda^{{}_{(XZ)}}_{{}_{(XY)}}}, \nonumber \\
         &= \frac{32}{7} \frac{DS}{2\ln(4/7)+\ln(D/2J)}.
\end{align}

\begin{figure}
  \begin{center}
    \includegraphics[width=\hsize]{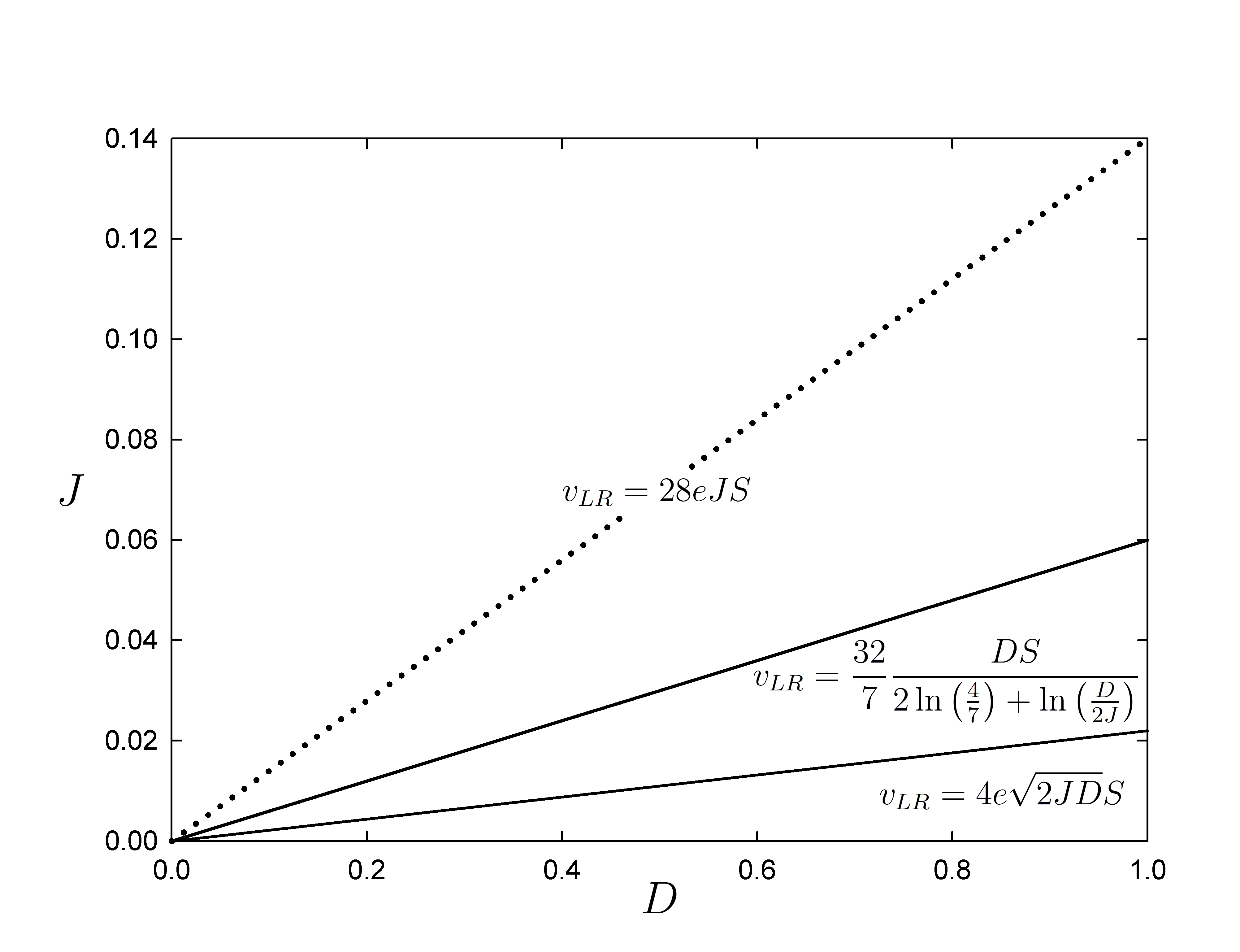}
  \end{center}
\caption{The (D,J)-plane of the XY-Model and the three different formulas for the Lieb-Robinson speed. If $ \frac{D}{2 J}\geq\left(\frac{7}{4}\right)^2 e^2 $, then the Lieb-Robinson speed is $4 e\sqrt{2 J D} S$; if $\left(\frac{7}{4}\right)^2 e\leq\frac{D}{2 J}\leq\left(\frac{7}{4}\right)^2 e^2$, then the Lieb-Robinson speed is $\frac{32}{7}\frac{D S}{2\ln(\frac{4}{7})+ \ln\left(\frac{D}{2 J}\right)} $; and finally, if $\frac{D}{2 J}\leq\left(\frac{7}{4}\right)^2 e$, the Lieb-Robinson speed is $28 e J S$. The dotted line indicates the critical line for the phase transition which runs along the line $\frac{D}{2 J} = 3.6$ }
\label{XYphase}
\end{figure}

The results and the various conditions are summarized in Table (\ref{vXY}).  The first column gives the Lieb-Robinson speed in the region of phase space defined by the inequalities of the third column.  By phase space, we simply mean the space $\R_+^2$ of all possible values of the coupling constants $J$ and $D$.  The different values of the Lieb-Robinson speed in the different regions of the phase space are plotted in Fig. \ref{XYphase}.  In general, the number of functional forms for the Lieb-Robinson speed could be as many as twice the number of distinct intersection points obtained in \rf{limbo} plus one, i.e. the minimum could occur at each intersection point or in each interval.

Also plotted in Fig. \ref{XYphase} is the critical line for the quantum phase transition (dotted line) which occurs for $\frac{D}{2 J} \approx 3.6$.  It is important to remember that the regions of phase space with different Lieb-Robinson speeds do not correspond to different quantum phases. Quantum phases occur in the ground state at a particular point of the phase space, whereas the Lieb-Robinson speed is the maximum speed of any signal in any state and not just small excitations over the ground state.

\begin{table}
\begin{tabular}{|c|c|c|}
\hline
 $v_{L R}$ & Conditions & Simplified Cond. \\
\hline
\hline
 \multirow{1}{*}{$v_{L R}^s(k_{X Z},\xi_{X Z}) $}  & \halfback$\lambda_{{}_{(X Z)}}^{{}_{(X Y)}} \geq \xi_{X Z}^{-1} $\halfback  &  $ \frac{D}{2 J}\geq\left(\frac{7}{4}\right)^2 e^2 $ \\
 $= 4 e\sqrt{2 J D} S$ & \halfback$\lambda_{{}_{(X Z)}}^{{}_{(X Y Z)}} \geq \xi_{X Z}^{-1}$\halfback & $ $\\
 & \halfback$\lambda_{{}_{(X Z)}}^{{}_{(X Y Z Y)}} \geq \xi_{X Z}^{-1}$\halfback & $ $\\
\hline
 $v_{L R}^d(k_{X Z},k_{X Y},$ &\halfback $\xi_{X Y}^{-1}\leq \lambda_{{}_{(X Z)}}^{{}_{(X Y )}} $\halfback & $ \left(\frac{7}{4}\right)^2 e\leq\frac{D}{2 J}\leq\left(\frac{7}{4}\right)^2 e^2 $   \\
  $\xi_{X Z},\xi_{X Y})$  & $ \lambda_{{}_{(X Z)}}^{{}_{(X Y )}}\leq  \xi_{X Z}^{-1}$ & \\
 &\halfback $ \lambda_{{}_{(X Y Z Y)}}^{{}_{(X Y)}}\!\!\leq\lambda_{{}_{(X Z)}}^{{}_{(X Y )}}$ \halfback  &  \\
$= \frac{32}{7}\frac{D S}{2\ln(\frac{4}{7})+ \ln\left(\frac{D}{2 J}\right)}$ & $ \lambda_{{}_{(X Y Z)}}^{{}_{(X Y)}}\leq\lambda_{{}_{(X Z)}}^{{}_{(X Y )}}$& \\
  \hline
 \multirow{1}{*}{$v_{L R}^s(k_{X Y},\xi_{X Y}) $}  &\halfback $\lambda_{{}_{(X Y Z)}}^{{}_{(X Y)}} \leq
 \xi_{X Y}^{-1}$\halfback  &  $  \frac{D}{2 J}\leq\left(\frac{7}{4}\right)^2 e $ \\
 $= 28 e J S$ &$\lambda_{{}_{(X Z)}}^{{}_{(X Y)}} \leq \xi_{X Y}^{-1}$ & $ $\\
 &$\lambda_{{}_{(X Y Z Y)}}^{{}_{(X Y)}} \leq \xi_{X Y}^{-1}$ & $ $\\
\hline

\end{tabular}
\caption{The Lieb-Robinson speed for the XY-model. The first column gives the functional form of the Lieb-Robinson speed in terms of the interaction strengths. The second column gives the region of applicability of each of the three functional forms of the Lieb-Robinson speed in terms of the $\lambda$'s and $\xi$'s. The last column gives the regions in terms conditions of the interaction strengths. The results of this table are depicted graphically in Fig. \ref{XYphase}.}\label{vXY}
\end{table}

\section*{Conclusions} \label{sec_conclusions}

In this article we showed how the Lieb-Robinson bound could be considerably tightened for the case of a finite number of interactions.  We also derived a bound which was sensitive to the relative interaction strengths.  We first considered the case of two interactions, which is one of the most frequently encountered cases.  Indeed, for a physically relevant bound we must use not the full Hamiltonian, but the effective Hamiltonian near the ground state, which often contains only two interactions.  We then generalised the result to any finite number of interactions. We observed that for $m\geq 3$ interactions, the space of coupling constants was subdivided into a finite number of regions, each of which produced a different functional form for the Lieb-Robinson speed. Interestingly, the dependence of the speed on the coupling is not necessarily algebraic anymore for three or more interactions as it was for two interactions; it was found that the logarithm of the coupling constants could appear in the speed.

It is also important, especially for applications in such domains as quantum gravity, that we be able to calculate the bound not only for a system on a lattice but also for general graphs. We derived explicit equations for the Lieb-Robinson bound and speed in the case of homogeneous and isotropic graphs. This is the best we can hope to have since if the graph is not homogeneous and isotropic, the speed will not be constant.

\appendix

\section*{Appendix A: Multiple Interactions Calculations} \label{appendix_multiple_interaction}
In \rf{trinductive} the end points of the chain are separated by a graph distance of $\sum_{k=1}^{n-1}D_{a(i_k)}^{a(i_k)\rightarrow a(i_{k+1})}$. Hence, if $d(P,Q)> \sum_{k=1}^{n-1}D_{a(i_k)}^{a(i_k)\rightarrow a(i_{k+1})}$, the chain cannot link $P$ to $Q$, where $a(i_k)\in\{1,2,\ldots,m\}$ labels the operator type of $i_k$ and $h(i_k)\in\{h_1,h_2, \ldots, h_m\}$ gives the coupling constant of operator $i_k$. We can thus legitimately multiply every term by $\exp\Big(\lambda \big(\sum_{k=1}^{n-1}D_{a(i_k)}^{a(i_k)\rightarrow a(i_{k+1})}- d(P,Q)\big)\Big)$ and drop the end-point condition. This gives us:
\begin{align}
\norm{{K_{a \ b}^{i_1 \ j}(t)}} & \leq  M \sum_{n=0}^\infty \frac{\abs{t}^n}{n!}\Big(\sum_{i_{2}\in Z_{i_1}}2 h(i_2)\ldots\sum_{i_{n+1}\in Z_{i_n}}2 h(i_{n+1})\Big) \exp\Big(\lambda \big(\sum_{k=1}^{n-1}D_{a(i_k)}^{a(i_k)\rightarrow a(i_{k+1})}- d(P,Q)\big)\Big) \nn
 & =  M e^{-\lambda d(P,Q)}\sum_{n=0}^\infty \frac{\abs{2 t}^n}{n!}\Big(\sum_{i_{2}\in Z_{i_1}} h(i_2)e^{\lambda D_{a(i_1)}^{a(i_1)\rightarrow a(i_{2})}}\ldots\sum_{i_{n+1}\in Z_{i_n}} h(i_{n+1})e^{\lambda D_{a(i_{n})}^{a(i_{n})\rightarrow a(i_{{n+1}})}}\Big) .
 \label{mothership}
 \end{align}

The equivalent equation for commutator bounded systems is:
\begin{align}
& \norm{{K_{a \ b}^{i_1 \ j}(t)}} \nn
& \leq  M \sum_{n=0}^\infty \frac{\abs{t}^n}{n!}\Big(\sum_{i_{2}\in Z_{i_1}}2 h(i_2)K_{a(i_1)a(i_2)}\ldots\sum_{i_{n+1}\in Z_{i_n}}2 K_{a(i_n)a(i_{n+1})}h(i_{n+1})\Big) \exp\Big(\lambda \big(\sum_{k=1}^{n-1}D_{a(i_k)}^{a(i_k)\rightarrow a(i_{k+1})}- d(P,Q)\big)\Big) \nn
 & =  M e^{-\lambda d(P,Q)}\sum_{n=0}^\infty \frac{\abs{2 t}^n}{n!}\Big(\sum_{i_{2}\in Z_{i_1}} K_{a(i_1)a(i_2)} h(i_2)e^{\lambda D_{a(i_1)}^{a(i_1)\rightarrow a(i_{2})}}\ldots\sum_{i_{n+1}\in Z_{i_n}} K_{a(i_n)a(i_{n+1})} h(i_{n+1})e^{\lambda D_{a(i_{n})}^{a(i_{n})\rightarrow a(i_{{n+1}})}}\Big) .
 \label{mothership22}
 \end{align}

For $(l_1,l_2)\in \{1,\ldots,m\}^2$, $l_1\neq l_2$ and $\lambda > 0 $, we define
\begin{align}
 N_{l_1 l_2}(\lambda) & \equiv n_{l_1\rightarrow l_2}h_{l_2} e^{\lambda D_{l_1}^{l_1\rightarrow l_2}} \text{for bounded systems} \nn
  N_{l_1 l_2}(\lambda) & \equiv n_{l_1\rightarrow l_2} K_{l_1 l_2} h_{l_2} e^{\lambda D_{l_1}^{l_1\rightarrow l_2}} \text{for commutator-bounded systems} \nn
 N_{l_1 l_1}(\lambda)& \equiv 0 \label{defN},
\end{align}

We proceed for bounded systems but note that the commutator-bounded derivation is virtually identical. Now, from an operator of type $1$, there are $n_{1\rightarrow l}$ ways to choose an operator of type $l$ as the next in the chain and there are $m-1$ possibilities for $l\in \{2,\ldots,m\}$, and similarly for the other operator types. So, if we consider the embedded sums
\begin{align}
 \sum_{i_{2}\in Z_{i_1}} h(i_2)e^{\lambda D_{a(i_1)}^{a(i_1)\rightarrow a(i_{2})}}\ldots\sum_{i_{n+1}\in Z_{i_n}} h(i_{n+1})e^{\lambda D_{a(i_{n})}^{a(i_{n})\rightarrow a(i_{{n+1}})}} ,\label{embsums}
\end{align}
we see that each individual sum of the type
\begin{align}
\sum_{i_{k+1}\in Z_{i_k}} h(i_{k+1})e^{\lambda D_{a(i_{k})}^{a(i_{k})\rightarrow a(i_{{k+1}})}}
\end{align}
can be written as
\begin{align}
 \sum_{l\neq a(i_k)} n_{a(i_{k})}^{a(i_{k})\rightarrow l} h_l e^{\lambda D_{a(i_{k})}^{a(i_{k})\rightarrow l}} = \sum_{l\neq a(i_k)} N_{a(i_{k})\rightarrow l}(\lambda) .
\end{align}
This means for fixed $n$ we can write
\begin{align}
 \sum_{i_{2}\in Z_{i_1}}\!\!\! h(i_2)e^{\lambda D_{a(i_1)}^{a(i_1)\rightarrow a(i_{2})}}\ldots\!\!\!\!\!\!\sum_{i_{n+1}\in Z_{i_n}} h(i_{n+1})e^{\lambda D_{a(i_{n})}^{a(i_{n})\rightarrow a(i_{{n+1}})}}
= \sum_{a_2\neq a(i_1)} N_{a(i_1) a_2}(\lambda)\ldots\sum_{a_{n+1}\neq a_n} N_{a_n a_{n+1}}(\lambda). \label{babyship}
\end{align}
of \rf{mothership}, where the $a_l$'s represent interaction types.  We thus have that \rf{babyship} has $(m-1)^n$ terms, each of which has the form
\begin{align}
N_{a_1 a_2}(\lambda) N_{a_2 a_3}(\lambda)N_{a_3 a_4}(\lambda) \ldots N_{a_n a_{n+1}}(\lambda) .\label{term}
\end{align}
Let us call a \emph{linking sequence of $N$'s of length $q$} a finite sequence of $N_{i j}(\lambda)$'s $N_{a_1 b_1}(\lambda), N_{a_2 b_2}(\lambda),\ldots, N_{a_q b_{q+1}}(\lambda)$ such that $a_{l+1}= b_l$.
Considering that there are only $m$ interactions, we have that the set of transition factors $\{N_{a b}(\lambda)| 1\leq a \leq m, 1\leq b \leq m, a\neq b\}$ contains $m(m-1)$ elements; thus, any sequence of $m(m-1)+1$ $N$'s will necessarily have at least one repetition. Hence, there are a finite number of linking sequences of $N$'s such that all $N$'s in the sequence are different, and they are of length $q\leq m(m-1)$. Let us call a linking sequence of $N$'s such that all $N$'s in the sequence are different and such that $a_1=b_{q+1}$ an \emph{elementary cycle}. To each elementary cycle $N_{a_1 a_2}(\lambda), N_{a_2 a_3}(\lambda),\ldots, N_{a_q a_{1}}(\lambda)$ we associate two numbers,
\begin{align}
k_{a_1 a_2 \ldots a_q} & \equiv \sqrt[q]{\prod_{l=1}^q h_{a_l}n_{a_l\rightarrow a_{l+1}}} \nn
\xi_{a_1 a_2 \ldots a_q} & \equiv  \frac{\sum_{l=1}^q D_{a_{l}}^{a_l\rightarrow a_{{l+1}}}}{q}\label{minicycl}
\end{align}
where we have made the identification $a_1=a_{q+1}$. We then have that
\begin{align}
N_{a_1 a_2}(\lambda) N_{a_2 a_3}(\lambda)\ldots N_{a_q a_{1}}(\lambda) = (k_{a_1 a_2 \ldots a_q}e^{\lambda \xi_{a_1 a_2 \ldots a_q}})^q. \label{kxiN}
\end{align}
Now \rf{term} is the product of a linking sequence of $N$'s of length $n$. In the first $m(m-1)+1$ $N$'s of the product there must be, as previously mentioned, at least two identical $N$'s; let us now call them $N_{b_1 b_2}$. Because we are dealing with the product of a linking sequence of $N$'s, the subsequence of $N$'s starting with the first $N_{b_1 b_2}$ and ending with $N$ just before the second $N_{b_1 b_2}$ is an elementary cycle of the type $N_{b_1 b_2}, N_{b_2 b_3},\ldots, N_{b_q b_{1}}$ and if we remove it from the full sequence, the remaining sequence is still a linking sequence of length $n-q$. So we can now write \rf{term} as
\begin{align}
(k_{b_1 b_2 \ldots b_q}e^{\lambda \xi_{b_1 b_2 \ldots b_q}})^q N_{c_1 c_2}(\lambda) \ldots N_{c_{n-q} c_{n+1-q}}(\lambda) ,  \label{term2}
\end{align}
where $N_{c_1 c_2}, \ldots, N_{c_{n-q} c_{n+1-q}}(\lambda)$ is a linking sequence of $N$'s. This process can be reiterated until at most $m(m-1)$ $N$'s are left. \\

\fbox{
\hspace{-1cm}\begin{minipage}[t]{0.48\textwidth}
\centering
\begin{align}
 & N_{1  3}\underbrace{N_{3  2}N_{2  4}N_{4  3}}_{(k_{3 2 4}e^{\lambda \xi_{3 2 4}})^3}N_{3 2}\underbrace{N_{2  4}N_{4 2}}_{(k_{2 4}e^{\lambda \xi_{2 4}})^2}N_{2 4}N_{4 3}N_{3 2}N_{2 1}N_{1 3}N_{3 4}\nn
 =  & (k_{3 2 4}e^{\lambda \xi_{3 2 4}})^3(k_{2 4}e^{\lambda \xi_{2 4}})^2N_{1  3}\underbrace{N_{3  2}N_{2 4}N_{4 3}}_{(k_{3 2 4}e^{\lambda \xi_{3 2 4}})^3}N_{3 2}N_{2 1}N_{1 3}N_{3 4}\nn
& =  (k_{3 2 4}e^{\lambda \xi_{3 2 4}})^6(k_{2 4}e^{\lambda \xi_{2 4}})^2\underbrace{N_{1  3}N_{3 2}N_{2 1}}_{(k_{1 3 2}e^{\lambda \xi_{1 3 2}})^3}N_{1 3}N_{3 4}\nn
& =  (k_{3 2 4}e^{\lambda \xi_{3 2 4}})^6 (k_{2 4}e^{\lambda \xi_{2 4}})^2 (k_{1 3 2}e^{\lambda \xi_{1 3 2}})^3N_{1 3}N_{3 4}.\label{exemple}
\end{align}
\begin{quote}
\small{Eq. (\ref{exemple})is an example of the rewriting of a term of the form in \rf{term} as a product of the elementary cycles defined in \rf{minicycl}. The dependence on $\lambda$ is not written explicitly in the above equation for simplicity.}
\end{quote}
\end{minipage}}
\\ \\

\noindent So if we now define
\begin{align}
 L(\lambda) & \equiv \max_{2\leq r\leq m(m-1)} \max_{(a_1, a_2, \ldots, a_r)\in \{1,\ldots,m\}^r} \bigg\{k_{a_1 \ldots a_r}e^{\lambda \xi_{a_1 a_r}}| 1\leq p<q\leq r, (a_p, a_{p+1})\neq (a_q,a_{q+1})\bigg\} \label{L}\\
 G(\lambda) & \equiv  \max_{(l_1,l_2,l_3,l_4)\in\{1,\ldots,m\}^4} \bigg\{\frac{N_{l_1 l_2}(\lambda)}{N_{l_3 l_4}(\lambda)}|N_{l_3 l_4}(\lambda)\neq 0\bigg\} < \infty , \label{G}
\end{align}
we conclude from the previous argument that
\begin{align}
N_{a_1 a_2}(\lambda) N_{a_2 a_3}(\lambda)N_{a_3 a_4}(\lambda) \ldots N_{a_n a_{n+1}}(\lambda)
\leq G(\lambda)^{m(m-1)}(L(\lambda))^n .\label{lessterm}
\end{align}
Since there are $(m-1)^n$ such terms in \rf{babyship}, this in turn gives us that
\begin{align}
 \sum_{i_{2}\in Z_{i_1}}\!\!\! h(i_2)e^{\lambda D_{a(i_1)}^{a(i_1)\rightarrow a(i_{2})}}\ldots\!\!\!\!\!\!\sum_{i_{n+1}\in Z_{i_n}} h(i_{n+1})e^{\lambda D_{a(i_{n})}^{a(i_{n})\rightarrow a(i_{{n+1}})}}
\leq G(\lambda)^{m(m-1)}\left((m-1) L(\lambda)\right)^n . \label{tricn}
\end{align}
Inserting this last inequality in \rf{mothership}, we obtain
\begin{align}
\norm{{K_{a \ b}^{i_1 \ j}(t)}} & \leq  \tilde{M}(\lambda) e^{-\lambda d(P,Q)}\sum_{n=0}^\infty \frac{\abs{2 t}^n}{n!} \big((m-1)L(\lambda)\big)^n \nn
 & =  \tilde{M}(\lambda)\exp \Big(\lambda \big( \frac{2(m-1)L(\lambda)}{\lambda}\abs{t} - d(P,Q)\big) \Big),
\end{align}
where $\tilde{M}(\lambda)= M (G(\lambda))^{M(m-1)}$. This last inequality implies a Lieb-Robinson speed of $\frac{2(m-1)L(\lambda)}{\lambda}$ for any positive $\lambda$. To obtain the tightest bound, we take the infimum over $\lambda$
\begin{align}
& \totalback v_{L R} =  \inf_{\lambda>0} \frac{2(m-1)L(\lambda)}{\lambda} .\label{vlr3}
\end{align}

\end{document}